\documentclass[12pt]{article}
\usepackage[english,german,french,polish]{babel}
\usepackage[T1]{fontenc}
\usepackage{amsfonts}

\begin{document}

\selectlanguage{english}

\baselineskip 0.75cm
\topmargin -0.6in
\oddsidemargin -0.1in

\let\ni=\noindent

\renewcommand{\thefootnote}{\fnsymbol{footnote}}

\pagestyle {plain}

\setcounter{page}{1}

\pagestyle{empty}

~~~

\begin{flushright}
IFT--03/3
\end{flushright}

{\large\centerline{\bf Bilarge mixing matrix and its invariance under}}

{\large\centerline{\bf "horizontal conjugation"~-- a new discrete transformation}}

{\large\centerline{\bf for neutrinos{\footnote {Work supported in part by the Polish State Committee for Scientific Research (KBN), grant 2 P03B 129 24 (2003--2004).}}}}

\vspace{0.4cm}

{\centerline {\sc Wojciech Kr\'{o}likowski}}

\vspace{0.3cm}

{\centerline {\it Institute of Theoretical Physics, Warsaw University}}

{\centerline {\it Ho\.{z}a 69,~~PL--00--681 Warszawa, ~Poland}}

\vspace{0.3cm}

{\centerline{\bf Abstract}}

\vspace{0.2cm}

In the first part of the note, we consider a neutrino texture, where the Dirac and right\-handed Majorana masses are proportional. If the former are approximately proportional 
also to the charged-lepton masses, then taking $\Delta m^2_{32} \sim 3\times 10^{-3}\;
{\rm eV}^2$ we estimate approximately that $\Delta m^2_{21} \sim O(10^{-5}\;
{\rm eV}^2)$, what is not very different from the recent KamLAND estimation $\Delta 
m^2_{21} \sim 7\times 10^{-5}\;{\rm eV}^2$, consistent with the LMA solar solution. In the second part, we show generically that the {\it invariance} of neutrino mixing matrix under 
the simultaneous discrete transformations $\nu_e \rightarrow -\nu_e $, $\nu_\mu 
\rightarrow \nu_\tau $, $\nu_\tau \rightarrow \nu_\mu $ and $\nu_1 \rightarrow -\nu_1 $, 
$\nu_2 \rightarrow -\nu_2 $, $\nu_3 \rightarrow \nu_3 $ (neutrino "horizontal conjugation") 
{\it characterizes} the familiar bilarge form of mixing matrix, favored phenomenologically 
at present. Then, in the case of this form, the mass neutrinos $\nu_1, \nu_2 , \nu_3 $ 
get a new quantum number, {\it covariant} in their mixings (neutrino "horizontal parity" \,equal to -1,-1,1, respectively). Conversely, such a covariance may be the {\it origin} of the bilarge mixing matrix. In Section 5, the "horizontal parity"~is  embedded in a group structure.

\vspace{0.2cm}

\ni PACS numbers: 12.15.Ff , 14.60.Pq , 12.15.Hh .

\vspace{0.6cm}

\ni March 2003  

\vfill\eject

~~~
\pagestyle {plain}

\setcounter{page}{1}

\vspace{0.2cm}

\ni {\bf 1. Introduction}

\vspace{0.3cm}

In a recent note [1] we considered the neutrino texture, where the Dirac and righthanded components $M^{(D)}$ and $M^{(R)}$ of the generic neutrino $6\times 6$ mass matrix

\vspace{-0.2cm}

\begin{equation}
\left( \begin{array}{cc} 0 & M^{(D)} \\ M^{(D)\,T} & M^{(R)} \end{array} \right) 
\end{equation}

\ni commute and have (at least, approximately) proportional eigenvalues,

\vspace{-0.2cm}

\begin{equation}
\lambda_1 {\bf :} \lambda_2 {\bf :} \lambda_3 = \Lambda_1 {\bf :} \Lambda_2 {\bf :} \Lambda_3 \,.
\end{equation}

\ni If, in addition, they are in a better or worse approximation proportional to the charged-lepton masses $m_e, m_\mu,\, m_\tau$, then in the seesaw mechanism the neutrino masses $m_1, m_2, m_3 $ being equal to $-\lambda_1^2/\Lambda_1, -\lambda_2^2/ \Lambda_2, -\lambda_3^2/\Lambda_3$, respectively, are approximately proportional to $m_e, m_\mu,\, m_\tau$. In fact,

\vspace{-0.1cm}

\begin{equation}
m_1 {\bf :} m_2 {\bf :} m_3 = \frac{\lambda_1^2}{\Lambda_1} {\bf :} \frac{\lambda_2^2}{\Lambda_2} {\bf :} \frac{\lambda_3^2}{\Lambda_3} = \lambda_1 {\bf :} \lambda_2 {\bf :} \lambda_3\simeq m_e {\bf :} m_\mu {\bf :} m_\tau .
\end{equation}

\ni In this case, we called the attention to the approximate relation

\vspace{-0.2cm}

\begin{equation}
\frac{\Delta m^2_{21}}{\Delta m^2_{32}} \simeq \frac{m^2_\mu - m^2_e}{m^2_\tau - m^2_\mu} =3.548\times 10^{-3}\,,
\end{equation}

\ni predicting the value

\vspace{-0.2cm}

\begin{equation}
\Delta m^2_{21} \sim 1.1\times 10^{-5}\;{\rm eV}^2\,,
\end{equation}

\ni when the SuperKamiokande~estimate $\Delta m^2_{32} \sim 3\times 10^{-3}\;{\rm eV}^2$ [2] is used (we assume that $\Delta m^2_{32} \geq 0$). Then,

\vspace{-0.2cm}

\begin{equation}
m^2_1 \sim 2.5\times 10^{-10}\;\,{\rm eV}^2\;,\; m^2_2 \sim 1.1\times 10^{-5}\;\,{\rm eV}^2 \;,\;m^2_3 \sim 3\times 10^{-3}\;\,{\rm eV}^2
\end{equation}

\ni and one obtains (with $m_1,m_2,m_3 >0$ and so, $\Lambda_1,\Lambda_2,\Lambda_3 < 0$)

\vspace{-0.2cm}

\begin{equation}
|\Lambda_1|\sim 1.7\times 10^{7}\; {\rm GeV}\,,\, |\Lambda_2| \sim 3.4\times 10^{9}\; {\rm GeV}\,,\, |\Lambda_3| \sim 5.8\times 10^{10}\; {\rm GeV}\,,
\end{equation}

\ni if one normalizes $\lambda_1^2 = m^2_e$ ({\it i.e.}, $|\Lambda_1| = \lambda_1^2/m_1 = m^2_e/m_1$). It is convenient to put $\lambda_1,\lambda_2,\lambda_3 > 0$. Of course, if the SuperKamiokande estimate for $\Delta m^2_{32}$ decreased, the prediction for $\Delta m^2_{21}$ would also decrease. 

The prediction (5) is not very different from the Large Mixing Angle MSW value supported confidently by the recent KamLAND experiment [3] and estimated as $ \Delta m^2_{21} \sim 7\times 10^{-5}\;{\rm eV}^2$ (the lower LMA solution) [4-8]. In order to get this value more precisely, one may put $ m_2/m_3 =\lambda_2/\lambda_3 \sim 2.6 m_\mu/m_\tau $ in place of $ m_2/m_3 =\lambda_2/\lambda_3 \simeq  m_\mu/m_\tau $, where $m_\tau/m_\mu = 16.82$ (here, Eq. (2) still  holds).

\vspace{0.3cm}

\ni {\bf 2. Dirac and Majorana masses: the conjecture of proportionality}

\vspace{0.3cm}

In the present note, we make the above considerations more operative putting (at least, approximately) $M^{(R)} = -\zeta M^{(D)}$ {\it i.e.}, $\Lambda_{1,2,3} = -\zeta \lambda_{1,2,3}$, where $\zeta \gg 1$ holds in consistency with the seesaw mechanism. In addition to this conjecture of proportionality, we will assume  the approximate proportionality $\lambda_1 {\bf :} \lambda_2 {\bf :} \lambda_3 \simeq m_e {\bf :} m_\mu {\bf :} m_\tau $, considered in Introduction, as normalized to the approximate equality $\lambda_{1,2,3} \simeq m_{e,\mu,\tau}$. Of course, the conjectured relation $M^{(R)}_{\alpha\,\beta} = -\zeta M^{(D)}_{\alpha\,\beta}\;({\alpha,\beta} = e,\mu,\tau)$ is valid after the spontaneous breaking of the electroweak symmetry $SU(2)_L \times U(1)_{Y}$ {\it i.e.}, when $M^{(D)}_{\alpha\,\beta} = Y^{(D)}_{\alpha\,\beta} \langle \phi^0 \rangle $. Then, also $M^{(R)}_{\alpha\,\beta}$ must include a mass scale. For instance, in the case of spontaneously broken left-right symmetry $SU(2)_L\times SU(2)_R\times U(1)_{B-L}$, where $Q = I^{(L)}_3 + Y/2$ and $Y/2 = I^{(R)}_3 + (B-L)/2$, this mass scale may be given by $\langle \phi^0_1 \rangle $ with $\vec{\phi}_1 = (\phi^{++}_1, \phi^+_1, \phi^0_1)$ denoting a Higgs right weak-isospin triplet (and a left weak-isospin singlet) which carries the (nonconserved) lepton number $L = -2$ (and the baryon number $B = 0$). In this case, $M^{(R)}_{\alpha\,\beta} = Y^{(R)}_{\alpha\,\beta} \langle \phi^0_1 \rangle = -\zeta Y^{(D)}_{\alpha\,\beta} \langle \phi^0 \rangle = -M^{(D)}_{\alpha\,\beta}$ according to our basic conjecture. If $Y^{(R)}_{\alpha\,\beta} = -Y^{(D)}_{\alpha\,\beta}$, then $ \langle \phi^0_1 \rangle = \zeta \langle \phi^0 \rangle >0 $.

Under these conjectures, the $6\times 6$ generic neutrino mass matrix (1) takes in the case of $M^{(D)T} = M^{(D)} = M^{(D)*}$ the form

\begin{equation}
\left( \begin{array}{cc} 0 & M^{(D)} \\ M^{(D)} & -\zeta M^{(D)} \end{array} \right) =
\left( \begin{array}{cc} 0 & {\bf 1}^{(3)} \\ {\bf 1}^{(3)} & -\zeta {\bf 1}^{(3)} \end{array} \right) 
\left( \begin{array}{cc} M^{(D)} & 0 \\ 0 & M^{(D)} \end{array} \right) \;,
\end{equation}

\vspace{0.1cm}

\ni where ${\bf 1}^{(3)} = {\rm diag}(1,1,1)$ and both matrix factors on the rhs commute. Using the diagonalizing matrix U for $M^{(D)}$ we obtain

\begin{eqnarray}
\lefteqn{\left( \begin{array}{cc} U^\dagger & 0 \\ 0 & U^\dagger \end{array} \right) 
\left( \begin{array}{cc} 0 & M^{(D)} \\ M^{(D)} & -\zeta M^{(D)} \end{array} \right) 
\left( \begin{array}{cc} U & 0 \\ 0 & U \end{array} \right)}\:\:\:\:\:\:\:\:\:\:\:\:\:\:\:\:\:\:\:\:\:\: \nonumber \\ & = & 
\left( \begin{array}{cc} 0 & {\bf 1}^{(3)} \\ {\bf 1}^{(3)} & -\zeta {\bf 1}^{(3)} \end{array} \right) 
\left( \begin{array}{cc} {\rm diag}\,(\lambda_1,  \lambda_2,  \lambda_3) & 0 \\ 0 & {\rm diag}\,(\lambda_1,  \lambda_2,  \lambda_3) \end{array} \right) \:,
\end{eqnarray}

\ni where both matrix factors on the rhs commute, of course. The diagonal form of the first matrix factor on the rhs of Eq. (9) turns out to be (for $\zeta >0$):

\begin{equation}
\left( \begin{array}{cc} \left[-\frac{\zeta}{2}+\sqrt{\left( \frac{\zeta}{2}\right)^2+1}\right]{\bf 1}^{(3)} & 0 \\ 0 & \left[-\frac{\zeta}{2}-\sqrt{\left( \frac{\zeta}{2} \right)^2+1}\right]{\bf 1}^{(3)} \end{array} \right) \,.
\end{equation}

\ni Thus, the diagonal form of the mass matrix (8) is (for $\zeta > 0$):

$$
\left( \begin{array}{cc} \left[-\frac{\zeta}{2}+\sqrt{\left( \frac{\zeta}{2}\right)^2+1}\right]\,{\rm diag}\,(\lambda_1,\lambda_2,\lambda_3) & 0 \\ 0 & \left[-\frac{\zeta}{2} -\sqrt{\left( \frac{\zeta}{2} \right)^2+1}\right]\,{\rm diag}\,(\lambda_1,  \lambda_2,  \lambda_3) \end{array} \right)
$$

\begin{equation}
 = \left( \begin{array}{cc} \frac{1}{\zeta}{\rm diag}\,(\lambda_1,\lambda_2,\lambda_3) & 0 \\ 0 & -\zeta {\rm diag}\,(\lambda_1,  \lambda_2,  \lambda_3) \end{array} \right) \,,
\end{equation}

\ni where the rhs is valid for $\zeta \gg 1$. Hence, for $\zeta \gg 1$

\begin{eqnarray}
m_{1,2,3} & = & \frac{1}{\zeta} \lambda_{1,2,3} \simeq \frac{1}{\zeta} m_{e,\mu,\tau} \simeq -\frac{1}{\zeta^2} m_{4,5,6} \,, \nonumber \\
m_{4,5,6} & = & -{\zeta} \lambda_{1,2,3} \simeq -{\zeta} m_{e,\mu,\tau} \simeq -{\zeta^2} m_{1,2,3}
\end{eqnarray}

\ni under our assumption of $\lambda_{1,2,3} \simeq m_{e,\mu,\tau}$. Using $m_\tau = 1776.99^{+0.29}_{-0.26}$ MeV and the SuperKamiokande estimate $\Delta m^2_{32} \sim 3\times 10^{-3} {\rm eV}^2$ giving $m_3 \sim 5.5\times 10^{-2}$ eV, we obtain 

\begin{equation}
\zeta = \frac{\lambda_3}{ m_{3} } \simeq \frac{m_\tau}{ m_{3} }  \sim 3.2\times 10^{10} 
\end{equation}

\ni and so, with $m_e = 0.510999$ MeV and $m_\mu = 105.658$ MeV we predict

\begin{equation}
 m_1  = \frac{1}{\zeta} \lambda_1 \simeq \frac{1}{\zeta} m_e \sim 1.6\times 10^{-5} \;{\rm eV}\;\,,\,\; m_2  = \frac{1}{\zeta} \lambda_2 \simeq \frac{1}{\zeta} m_\mu \sim 3.3\times 10^{-3} \;{\rm eV}\,.
\end{equation}

\ni Thus, $m^2_1 \sim 2.5\times 10^{-10} \;{\rm eV}^2$, $m^2_2 \sim 1.1\times 10^{-5}\; {\rm eV}^2$ and $m^2_3  \sim 3\times 10^{-3} \;{\rm eV}^2$.

The KamLAND estimate $\Delta m^2_{21} \sim 7\times 10^{-5} \;{\rm eV}^2$ gives the value $ m_2  \sim 8.4\times 10^{-3}$ eV which lies not so far from our parameterfree prediction $ m_2  \sim 3.3\times 10^{-3}$ eV. Note that we may get the KamLAND value more precisely putting $ m_2/m_3 = \lambda_2/\lambda_3 \sim 2.6 m_\mu /m_\tau$ {\it i.e.}, for instance, $\zeta  m_1  = \lambda_1 \sim m_e$, $\zeta  m_2  = \lambda_2 \sim 0.90 m_\mu$ and $\zeta  m_3  = \lambda_3 \sim 0.35 m_\tau$ in place of $\zeta  m_2  = \lambda_2 \simeq  m_\mu$ and $\zeta  m_3  = \lambda_3 \simeq m_\tau$, as 0.90/0.35 = 2.6 (here, Eq. (2) still holds, now with $\Lambda_{1,2,3} = -\zeta \lambda_{1,2,3}$). In this example, $\zeta = \lambda_3/ m_3  \sim 0.35 m_\tau / m_3  \sim 1.1\times 10^{10}$. 

It is natural that the neutrino Dirac masses $\lambda_1,\lambda_2,\lambda_3 $ may be not equal to the charged-lepton masses $ m_e , m_\mu , m_\tau $ (in fact, somewhat smaller than these masses because of electromagnetic interactions of charged leptons). Of course, the neutrino Majorana masses $ m_1 , m_2 , m_3 $ are dramatically smaller than $ m_e , m_\mu , m_\tau $, as $ m_{1,2,3}  \simeq \lambda_{1,2,3}/\zeta \ll \lambda_{1,2,3} \leq m_{e,\mu,\tau}$ due to $\zeta \gg 1$. It seems also natural that the lepton Dirac masses $\lambda_1,\lambda_2,\lambda_3 $ and $ m_e , m_\mu , m_\tau $ are smaller than the masses of respective up and down quarks, since quarks participate in strong interactions. 

\vspace{0.3cm}

\ni {\bf 3. Bilarge mixing matrix: the invariance induced by $\nu_1,\nu_2,\nu_3 \rightarrow -\nu_1,-\nu_2,\nu_3 $}

\vspace{0.3cm}

In our texture, where 

\vspace{-0.3cm}

\begin{equation}
M^{\rm eff} \equiv \left[-\frac{\zeta}{2}+\sqrt{\left( \frac{\zeta}{2} \right)^2+1}\right] M^{(D)} = \frac{1}{\zeta} M^{(D)}
\end{equation}

\ni ($\zeta \gg 1$) with

\vspace{-0.3cm}

\begin{equation}
U^\dagger M^{(D)} U = {\rm diag} (\lambda_1, \lambda_2, \lambda_3) \,,
\end{equation}

\ni and in consequence

\vspace{-0.2cm}

\begin{equation}
m_{1,2,3}  = \left[-\frac{\zeta}{2}+\sqrt{\left( \frac{\zeta}{2} \right)^2+1}\right] \lambda_{1,2,3} = \frac{1}{\zeta}\lambda_{1,2,3}  
\end{equation}

\ni ($\zeta \gg 1$),  the form of the Dirac mass matrix $M^{(D)}$ is unknown. In the situation, when the form of effective mass matrix $M^{\rm eff}$ for active neutrinos $\nu_e, \nu_\mu, \nu_\tau $ is  theoretically {\it not known} enough, the questions of the neutrino mass spectrum $m_1, m_2, m_3$ and of the diagonalizing matrix $U$ for $M^{\rm eff}$,

\begin{equation}
U^\dagger M^{\rm eff} U = {\rm diag} (m_1, m_2, m_3) \,,
\end{equation}

\ni are phenomenologically {\it independent}, though they lead jointly to 

\vspace{-0.2cm}

\begin{equation}
M^{\rm eff} = U {\rm diag} (m_1, m_2, m_3) U^\dagger \,.
\end{equation}

\ni This independence enables, {\it a priori}, a hierarchical mass spectrum to coexist with a large mixing of neutrino states by the diagonalizing matrix.

In the flavor representation, where the mass matrix for charged leptons is diagonal, the neutrino diagonalizing matrix $ U = (U_{\alpha i})$ ($\alpha = e,\mu,\tau$ and $i = 1,2,3$) is at the same time the mixing matrix for active neutrinos according to the unitary transformation

\vspace{-0.3cm}

\begin{equation}
\nu_{\alpha}  = \sum_i U_{\alpha i}  \nu_i \;,
\end{equation}

\ni where $\nu_\alpha \equiv \nu_{\alpha L}$ and $\nu_i \equiv \nu_{i L}$ denote the active-neutrino flavor and mass fields, respectively. As is well known, the bilarge form of the mixing matrix

\begin{equation}
U = \left( \begin{array}{ccc} c_{12} & s_{12} & 0 \\ - \frac{1}{\sqrt2} s_{12} & \frac{1}{\sqrt2} c_{12} & \frac{1}{\sqrt2}  \\ \frac{1}{\sqrt2} s_{12} & -\frac{1}{\sqrt2} c_{12} & \frac{1}{\sqrt2}  \end{array} \right)\;, 
\end{equation}

\ni where $c_{23} = 1/\sqrt2\ = s_{23}$ and $s_{13} = 0$ (and $s_{12} < c_{12}$ with $\theta_{12} \sim 33^\circ$ [4-8] are also large), is globally consistent with all present neutrino oscillation experiments for solar $\nu_e$'s and atmospheric $\nu_\mu$'s as well as with the negative Chooz experiment [9] for reactor $\bar\nu_e$'s (giving $s^2_{13} < 0.03$), but it cannot explain the possible LSND effect [10] for accelerator $\bar\nu_\mu$'s (and $\nu_\mu$'s) whose existence is expected to be clarified soon in the MiniBOONE experiment (in Ref. [11] a "$\!$\,last hope"\, for explaining the possible LSND effect by a hypothetic sterile neutrino is considered). 

In the case of the mixing matrix $U$ as given in Eq. (21), the unitary transformation (20) gets the form

\vspace{-0.2cm}

\begin{eqnarray}
\nu_e & = & \;\;\;c_{12} \nu_1 + s_{12} \nu_2 \,, \nonumber \\ \nu_\mu & = & -\frac{1}{\sqrt2}(s_{12} \nu_1 - c_{12} \nu_2) + \frac{1}{\sqrt2} \nu_3 \,, \nonumber \\ \nu_\tau & = & \;\;\; \frac{1}{\sqrt2}(s_{12} \nu_1 - c_{12} \nu_2) + \frac{1}{\sqrt2} \nu_3 \,,
\end{eqnarray}

\ni while the inverse transformation reads

\vspace{-0.2cm}

\begin{eqnarray}
\nu_1 & = & c_{12} \nu_e - s_{12}\frac{1}{\sqrt2} (\nu_\mu - \nu_\tau) \,, \nonumber \\ \nu_2 & = & s_{12} \nu_e + c_{12}\frac{1}{\sqrt2} (\nu_\mu - \nu_\tau) \,, \nonumber \\ \nu_3 & = & \frac{1}{\sqrt2} (\nu_\mu + \nu_\tau) \,.
\end{eqnarray}

\ni It can be seen that due to Eq. (22) the discrete transformation $\nu_1 \rightarrow -\nu_1 $, $\nu_2 \rightarrow -\nu_2 $, $\nu_3 \rightarrow \nu_3$ of mass neutrinos induces for flavor neutrinos the discrete transformation $\nu_e \rightarrow -\nu_e $, $\nu_\mu \rightarrow \nu_\tau $, $\nu_\tau \rightarrow \nu_\mu $ {\it i.e.}, the change of sign of $\nu_e$ and the interchange of $\nu_\mu$ and $\nu_\tau$ [this is a consequence of the maximal mixing of $\nu_\mu$ and $\nu_\tau$ in Eqs. (23)]. We can conclude that the above interplay between both discrete transformations {\it characterizes} the form (21) of mixing matrix and so, if conjectured, {\it selects} such a form (for any $c_{12}$ and $s_{12}$) from its other possible forms. Formally, we infer that the above interplay is realized just in the case of $U$ given in Eq. (21) because of the relations

\begin{eqnarray}
\left( \begin{array}{r} -\nu_1 \\ -\nu_2 \\ \nu_3 \end{array} \right) & = & 
\left( \begin{array}{rrr} -1& 0 & 0 \\ 0 & -1 & 0 \\ 0 & 0 & 1 \end{array} \right) \left( \begin{array}{r} \nu_1 \\ \nu_2 \\ \nu_3 \end{array} \right) \;,\; \nonumber \\ 
\left( \begin{array}{r} -\nu_e \\ \nu_\tau \\ \nu_\mu \end{array} \right) & = &  
\left( \begin{array}{rrr} -1& 0 & 0 \\ 0 & 0 & 1 \\ 0 & 1 & 0 \end{array} \right) \left( \begin{array}{r} \nu_e \\ \nu_\mu \\ \nu_\tau \end{array} \right)
\end{eqnarray}

\ni and

\begin{equation}
\left( \begin{array}{rrr} -1 & 0 & 0 \\ 0 & 0 & 1\\ 0 & 1 & 0 \end{array} \right) U \left( \begin{array}{rrr} -1 & 0 & 0 \\ 0 & -1 & 0 \\ 0 & 0 & 1 \end{array}\right) = U \;, 
\end{equation}

\ni where due to Eq. (20)

\begin{equation}
\left( \begin{array}{r} \nu_e \\ \nu_\mu \\ \nu_\tau \end{array} \right) = U \left( \begin{array}{r} \nu_1 \\ \nu_2 \\ \nu_3 \end{array} \right) \;.
\end{equation}

\ni Here, the relation (25) is crucial, telling us that the mixing matrix $U$ of the form (21)  {\it is invariant} under the simultaneous transformations $\nu_e \rightarrow -\nu_e $, $\nu_\mu \rightarrow \nu_\tau $, $\nu_\tau \rightarrow \nu_\mu $ and $\nu_1 \rightarrow -\nu_1 $, $\nu_2 \rightarrow -\nu_2 $, $\nu_3 \rightarrow \nu_3 $.  Given such a mixing matrix $U$, the first transformation is induced by the second through the unitary transformation between their matrices: $(I) = U (II) U^\dagger$. That is equivalent to $(I) U (II) = U$ {\it i.e.}, to Eq. (25).

Making use of the (formal) horizontal $SU(3)$ group generated by ${\widehat{\lambda}_a}/2 \;(a = 1,2,\ldots,8)$ with $\widehat{\lambda}_a$ being the Gell-Mann $3\times 3$ matrices acting on the horizontal triplet $(\nu_1, \nu_2, \nu_3)^T$, we can realize the above discrete transformations for mass and flavor neutrinos by means of the matrices

\begin{eqnarray}
\left(\!\! \begin{array}{rrr}  -1\! & \!0 & 0 \\ 0\! & \!-1 & 0 \\ 0\! & \!0 & 1 \end{array}\right) & = & -\frac{1}{3}\widehat{1} - \frac{2}{\sqrt3} \widehat{\lambda}_8 \;, \nonumber \\
\left( \begin{array}{rrr} \!-1 & 0 & 0 \\ 0 & 0 &1 \\ 0 & 1 & 0 \end{array} \right) & = & -\frac{1}{3} \widehat{1} - \frac{1}{2}\left(\widehat{\lambda}_3 + \frac{1}{\sqrt3}\widehat{\lambda}_8 \right) + \widehat{\lambda}_6 \,,
\end{eqnarray}

\ni where 

\begin{equation}
\widehat{1} = \left( \begin{array}{rrr} 1 & 0 & 0 \\ 0 & 1 & 0  \\ 0 & 0 & 1 \end{array} \right) ,\,\widehat{\lambda}_3 = \left( \begin{array}{rrr} 1 & 0 & 0 \\ 0 & -1 & 0\\ 0 & 0 & 0 \end{array}\right) ,\, \widehat{\lambda}_8 =  \frac{1}{\sqrt3} \left( \begin{array}{rrr} 1 & 0 & 0 \\ 0 & 1 & 0 \\ 0 & 0 & -2 \end{array}\right) ,\, \widehat{\lambda}_6 = \left( \begin{array}{rrr} 0 & 0 & 0 \\ 0 & 0 & 1\\ 0 & 1 & 0 \end{array}\right) . 
\end{equation}

\ni Note that

\begin{equation}
Q^{(H)} \equiv \frac{1}{2}\left(\widehat{\lambda}_3 + \frac{1}{\sqrt3}\widehat{\lambda}_8 \right) = \left( \begin{array}{ccc} 2/3 & 0 & 0 \\ 0 & -1/3\; & 0 \\ 0 & 0 & -1/3\, \end{array} \right) 
\end{equation}

\ni plays a role of "horizontal charge", while $ I^{(H)}_3 \equiv \widehat{\lambda}_3/2$ and $Y^{(H)} \equiv \widehat{\lambda}_8/{\sqrt3}$ are the 3-component of the "horizontal isospin"~and the "horizontal hypercharge", respectively.

Notice also that the mixing matrix (21) can be written in the form

\begin{equation}
U = e^{i\widehat{\lambda}_7 \,\pi/4} e^{i\widehat{\lambda}_2 \,\theta_{12}} \,,
\end{equation}

\ni since

\begin{equation}
e^{i\widehat{\lambda}_7 \theta_{23}} = \left( \begin{array}{rcc} 1 & 0 & 0 \\ 0 & c_{23} & s_{23} \\ 0 & -s_{23}\;\; & c_{23} \end{array} \right)\,,\, \widehat{\lambda}_7 = \left( \begin{array}{rrr} 0 & 0 & 0 \\ 0 & 0 & -i \\ 0 & i & 0 \end{array} \right) 
\end{equation}

\ni and

\begin{equation}
e^{i\widehat{\lambda}_2 \theta_{12}} = \left( \begin{array}{ccr}  c_{12} & s_{12} & 0 \\ -s_{12}\;\;  & c_{12} & 0 \\ 0 & 0 & 1  \end{array} \right)\,,\, \widehat{\lambda}_2 = \left( \begin{array}{rrr} 0 & -i & 0 \\ i & 0 & 0 \\ 0 & 0 & 0 \end{array} \right) \,,
\end{equation}

\ni where $\theta_{23} = \pi/4 $.

We can see from Eq. (32) that

\begin{equation}
e^{i\widehat{\lambda}_2 \pi} = \left( \begin{array}{rrr} -1 & 0 & 0 \\ 0 & -1 & 0  \\ 0 & 0 & 1 \end{array} \right)  = \left( \begin{array}{rrr} -1 & 0 & 0 \\ 0 & -1 & 0  \\ 0 & 0 & 1 \end{array} \right)^{-1} 
\end{equation}

\ni and so, our discrete transformation of mass neutrinos can be realized as

\begin{equation}
e^{i\widehat{\lambda}_2 \pi} \left( \begin{array}{r} \nu_1 \\ \nu_2 \\ \nu_3 \end{array} \right) = \left( \begin{array}{r} -\nu_1 \\ -\nu_2 \\ \nu_3 \end{array} \right) 
\end{equation}

\ni {\it i.e.}, as the unitary rotation around the "horizontal"~2-axis by angle $2\pi $, generated by the~2-component $ \!I^{(H)}_2 \!\equiv \!{\widehat{\lambda}_2}/2$ of the "horizontal isospin". The discrete transformation $\nu_1 \rightarrow -\nu_1 $, $\nu_2 \rightarrow -\nu_2 $, $\nu_3 \rightarrow \nu_3 $ realized in Eq. (34) may be called the "horizontal conjugation", while its matrix

\begin{equation}
P^{(H)} \equiv e^{i I^{(H)}_2\, 2\pi} = e^{i\widehat{\lambda}_2\, \pi}
\end{equation}

\ni given in Eq. (33) --- the "horizontal parity". Then, the mass neutrinos $\nu_1, \nu_2, \nu_3 $ correspond to the eigenvalues -1,-1,1 of this parity, respectively; also the flavor neutrino $\nu_e $ gets the eigenvalue -1, while $\nu_\mu $ and $ \nu_\tau $ mix the eigenvalues -1 and 1 in such a way that $(\nu_\mu \mp \nu_\tau)/{\sqrt2}$ have the eigenvalues $\mp 1$. Thus, it follows from the unitary transformation (23) that for the form (21) of mixing matrix $U$ the "horizontal parity"~(35) is an observable {\it covariant} in neutrino mixings: ${P^{(H)}}' = U P^{(H)} U^\dagger$. This is equivalent to ${P^{(H)}}' U P^{(H)} = U$ {\it i.e.}, to Eq. (25).

At the same time, we can infer from the relation (25) and Eqs. (30) with (33) that

\begin{equation}
e^{i\widehat{\lambda}_7 \pi/4} e^{i\widehat{\lambda}_2 \pi} e^{-i\widehat{\lambda}_7 \pi/4} = \left( \begin{array}{rrr} \!-1 & 0 & 0 \\ 0 & 0 &1 \\ 0 & 1 & 0 \end{array} \right) = \left( \begin{array}{rrr} \!-1 & 0 & 0 \\ 0 & 0 &1 \\ 0 & 1 & 0 \end{array} \right)^{-1} 
\end{equation}

\ni and thus, in consequence of the unitary rotation (34) for mass neutrinos, the composed unitary rotation

\begin{equation}
e^{i\widehat{\lambda}_7 \pi/4} e^{i\widehat{\lambda}_2 \pi} e^{-i\widehat{\lambda}_7 \pi/4} \left( \begin{array}{r} \nu_e \\ \nu_\mu \\ \nu_\tau \end{array} \right) = \left( \begin{array}{r} -\nu_e \\ \nu_\tau \\ \nu_\mu \end{array} \right)
\end{equation}

\ni is induced for flavour neutrinos. This interplay between the discrete transformations (34)
and (37) selects the form (21) of mixing matrix $U$ (for any $c_{12}$ and $s_{12}$) as satisfying the relation (25) that now is identically fulfilled, being reduced trivially to

\begin{equation}
e^{i\widehat{\lambda}_7 \pi/4} e^{i\widehat{\lambda}_2 \theta_{12}} = e^{i\widehat{\lambda}_7 \pi/4} e^{i\widehat{\lambda}_2 \theta_{12}}
\end{equation}

\ni due to Eqs. (30), (33) and (36).

\vspace{0.3cm}

\ni {\bf 4. Effective mass matrix: the invariance under $\nu_e,\nu_\mu,\nu_\tau \rightarrow -\nu_e, \nu_\tau,\nu_\mu $}

\vspace{0.3cm}

Making use of the formulae (19) and (21), and parametrizing the neutrino mass spectrum as

\begin{equation}
m_1 = \stackrel{0}{m} - \delta\,,\, m_2 = \stackrel{0}{m} + \delta\,,\, m_3 = \stackrel{0}{m} + \Delta\,,
\end{equation}

\ni we can write the effective mass matrix for active neutrinos $\nu_e, \nu_\mu, \nu_\tau$ in the form

\begin{equation}
M^{\rm eff} = \stackrel{0}{m} \left( \begin{array}{rrr} 1 & 0 & 0 \\ 0 & 1 & 0 \\ 0 & 0 &1 \end{array} \right) + \frac{1}{2} \Delta \left( \begin{array}{rrr} 0 & 0 & 0 \\ 0 & 1 & 1 \\ 0 & 1 & 1 \end{array}\right) + \frac{1}{2} \delta \left( \begin{array}{rrr} -2c & \sqrt{2}\,s & -\sqrt{2}\,s \\ \sqrt{2}\,s & c & -c \\ -\sqrt{2}\,s & -c & c \end{array}\right) \,,
\end{equation}

\ni where $c \equiv \cos 2\theta_{12} = c_{12}^2 - s_{12}^2$ and $s \equiv \sin 2\theta_{12} =2c_{12} s_{12}$. In Eq. (40) all three terms commute (the product of the second and third term in both orders vanishes). Thus, consistently

\begin{equation}
{\rm diag} (m_1,m_2,m_3) = U^\dagger M^{\rm eff}U  
 = \stackrel{0}{m} \left( \begin{array}{rrr} 1 & 0 & 0 \\ 0 & 1 & 0 \\ 0 & 0 &1 \end{array} \right) + \Delta \left( \begin{array}{rrr} 0 & 0 & 0 \\ 0 & 0 & 0 \\ 0 & 0 & 1 \end{array} \right) + \delta \left( \begin{array}{rrr} -1 & 0 & 0 \\ 0 & 1 & 0 \\ 0 & 0 & 0 \end{array} \right) \,. 
\end{equation}

\ni In Eq. (40), the third term is equal to the sum

\begin{equation}
\frac{1}{2} \delta \left( \begin{array}{rrr} -2 & 0 & 0 \\ 0 & 1 & -1 \\ 0 & -1 & 1 \end{array} \right) c + \frac{1}{\sqrt2}\delta \left( \begin{array}{rrr} 0 & 1 & -1 \\ 1 & 0 & 0 \\ -1 & 0 & 0 \end{array} \right) s\,.
\end{equation}

\ni Formally, in deriving Eq. (41) from Eq. (40) the relations

\begin{eqnarray}
U^\dagger \left( \begin{array}{rrr} 0 & 0 & 0 \\ 0 & 1 & 1 \\ 0 & 1 & 1 \end{array} \right) U & =  & 2 \left( \begin{array}{rrr} 0 & 0 & 0 \\ 0 & 0 & 0 \\ 0 & 0 & 1 \end{array}\right) \,, \nonumber \\ & & \nonumber \\ U^\dagger \left( \begin{array}{rrr} -2 & 0 & 0 \\ 0 & 1 & -1 \\ 0 & -1 & 1 \end{array} \right) U & =  & 2 \left( \begin{array}{rrr} -c & -s & 0 \\ -s & c & 0 \\ 0 & 0 & 0 \end{array}\right) \,, \nonumber \\ & & \nonumber \\ U^\dagger \left( \begin{array}{rrr} 0 & 1 & -1 \\ 1 & 0 & 0 \\ -1 & 0 & 0 \end{array} \right) U & =  & \!\!\!\!\! \sqrt2 \left( \begin{array}{rrr} -s & c & 0 \\ c & s & 0 \\ 0 & 0 & 0 \end{array}\right)
\end{eqnarray}

\ni are involved.

We can easily check that the effective mass matrix (40) of flavor neutrinos $\nu_e, \nu_\mu, \nu_\tau$ {\it is invariant} under the discrete transformation $\nu_e \rightarrow -\nu_e $, $\nu_\mu 
\rightarrow \nu_\tau $, $\nu_\tau \rightarrow \nu_\mu $ induced by $\nu_1 \rightarrow -\nu_1 $, $\nu_2 \rightarrow -\nu_2 $, $\nu_3 \rightarrow \nu_3 $ ("horizontal conjugation"):

\begin{equation}
\left( \begin{array}{rrr}  -1 & 0 & 0 \\ 0 & 0 & 1 \\ 0 & 1 & 0 \end{array}\right) M^{\rm eff} \left(  \begin{array}{rrr}  -1 & 0 & 0 \\ 0 & 0 & 1 \\ 0 & 1 & 0 \end{array}\right) = M^{\rm eff} \,.
\end{equation}

\ni In fact, all matrices appearing in Eqs. (40) and (42) commute with the transformation matrix in Eq. (44) that squared gives the unit matrix. The invariance (44) follows also directly from the formula (19) and the relation (25).

In terms of the Gell-Mann $3\times 3$ matrices $\widehat{\lambda}_a \;(a = 1,2,\ldots,8)$ we can put in Eqs. (40) and (42)

\begin{eqnarray}
\left(\; \begin{array}{rrrrr}  0  &  \! & 0  &  \! & 0 \\ 0  &  \! & 1  &  \! & 1 \\ 0  &  \! & 1  &  \! & 1 \end{array}\right) & = & \frac{2}{3}\widehat{1} - \frac{1}{2} (\widehat{\lambda}_3 + \frac{1}{\sqrt3} \widehat{\lambda}_8) + \widehat{\lambda}_6 \;, \nonumber \\
\left(\!\!\! \begin{array}{rrr}  -2 & 0 & 0 \\ 0 & 1 & -1 \\ 0 & -1 & 1 \end{array}\right) & = & -\frac{3}{2}(\widehat{\lambda}_3 + \frac{1}{\sqrt3} \widehat{\lambda}_8) - \widehat{\lambda}_6 \;, \nonumber \\
\left(\!\!\! \begin{array}{rrrr} 0  &  \! & 1 & -1 \\ 1  &  \! & 0 & 0 \\ -1  &  \! & 0 & 0 \end{array} \right) & = &  \widehat{\lambda}_1 - \widehat{\lambda}_4 \,,
\end{eqnarray}

\ni where in addition to Eqs. (28) we use

\begin{equation}
\widehat{\lambda}_1 = \left( \begin{array}{rrr} 0 & 1 & 0 \\ 1 & 0 & 0\\ 0 & 0 & 0 \end{array}\right) ,\, \widehat{\lambda}_4 = \left( \begin{array}{rrr} 0 & 0 & 1 \\ 0 & 0 & 0 \\ 1 & 0 & 0  \end{array}\right) \,.
\end{equation}

\ni Then, from Eqs. (40) and (42) we obtain the formula 

\begin{equation}
M^{\rm eff} = \left(\stackrel{0}{m}+ \frac{1}{3} \Delta \right) \widehat{1} - \frac{1}{2}\left(\Delta + 3\delta c \right) \frac{1}{2} \left( \widehat{\lambda}_3 + \frac{1}{\sqrt3} \widehat{\lambda}_8 \right) + \frac{1}{2} \left(\Delta - \delta c \right) \widehat{\lambda}_6 + \frac{1}{\sqrt2} \delta s \left(\widehat{\lambda}_1 - \widehat{\lambda}_4 \right)\,.
\end{equation}

\vspace{0.2cm}

The effective mass matrix $ M^{\rm eff} = \left(M_{\alpha\,\beta} \right)\;(\alpha,\beta = e,\mu,\tau)$ may be also presented as

\begin{equation}
M^{\rm eff} = \sum_{\alpha\,\beta} M_{\alpha \beta}\, \widehat{e}_{\alpha\,\beta}
\end{equation}

\ni in terms of the basic matrices $\widehat{e}_{\alpha\,\beta} \equiv (\delta_{\alpha\,\gamma}\, \delta_{\beta\,\delta}$), where

\begin{eqnarray}
\widehat{e}_{e e} = \frac{1}{3}\widehat{1} + \frac{1}{2}\left(\,\; \widehat{\lambda}_3 + \frac{1}{\sqrt3} \widehat{\lambda}_8 \right)  & , & \widehat{e}_{e \mu} = \frac{1}{2}\left(\widehat{\lambda}_1 + i \widehat{\lambda}_2 \right) = \widehat{e}^\dagger_{\mu e}\;\;, \nonumber \\ \widehat{e}_{\mu \mu} = \frac{1}{3}\widehat{1} + \frac{1}{2} \!\left(\!\!-\widehat{\lambda}_3 + \frac{1}{\sqrt3} \widehat{\lambda}_8 \right)  & , & \widehat{e}_{e \tau} = \frac{1}{2}\left(\widehat{\lambda}_4 + i \widehat{\lambda}_5 \right) = \widehat{e}^\dagger_{\tau e} \;\;, \nonumber \\ \widehat{e}_{\tau \tau} = \frac{1}{3}\widehat{1} - \frac{1}{\sqrt3} \widehat{\lambda}_8 \;\;\;\;\;\;\;\;\;\;\;\;\;\;\;\;\; & , & \widehat{e}_{\mu \tau} = \frac{1}{2}\left(\widehat{\lambda}_6 + i \widehat{\lambda}_7 \right) = \widehat{e}^\dagger_{\tau \mu}  
\end{eqnarray}

\vspace{0.2cm}

\ni with $\widehat{1} = (\delta_{\gamma\,\delta})$ and $\widehat{\lambda}_a = (\lambda_{a\, \gamma \delta})\;(a = 1,2,\ldots,8)$. The matrix elements $M_{\alpha \beta}$ of $M^{\rm eff}$ are determined by the formula (19), $M_{\alpha \beta} = \sum_i U_{\alpha i} m_i U^*_{\beta i}$, that due to Eqs. (21) and(39) gives

\begin{eqnarray}
M_{e e} & = & \;\,\stackrel{0}{m} - \delta c \,, \nonumber \\ M_{\mu \mu} & = & \;\,M_{\tau \tau} = \;\stackrel{0}{m} + \frac{1}{2}\Delta + \frac{1}{2}\delta c \,, \nonumber \\ 
M_{e \mu} & = & \! -M_{e \tau} = \frac{1}{\sqrt2}\delta s = M_{\mu e} = -M_{\tau e} \,, \nonumber \\ 
M_{\mu \tau} & = & \;\,\frac{1}{2}\Delta - \frac{1}{2}\delta c  = -M_{\tau \mu} \,.
\end{eqnarray}

\vspace{0.2cm}

\ni It may be worthwhile to note that the imaginary matrices $\widehat{\lambda}_2, \widehat{\lambda}_7$ and

\begin{equation}
\widehat{\lambda}_5 = \left( \begin{array}{rrr} 0 & 0 & -i \\ 0 & 0 & 0 \\ i & 0 & 0  \end{array}\right) \,,
\end{equation}

\vspace{0.2cm}

\ni although they appear within the basic matrices $\widehat{e}_{\alpha \beta}$, are cancelled out in $M^{\rm eff}$ due to the relations $ M_{\beta \alpha} = M_{\alpha \beta}$ and $\widehat{e}_{\beta \alpha} = \widehat{e}^\dagger_{\alpha \beta}$ applied to Eq. (48).

Finally, it is tempting to speculate that the invariance under the simultaneous discrete transformations $\nu_e \rightarrow -\nu_e $, $\nu_\mu \rightarrow \nu_\tau $, $\nu_\tau \rightarrow \nu_\mu $ and $\nu_1 \rightarrow -\nu_1 $, $\nu_2 \rightarrow -\nu_2 $, $\nu_3 \rightarrow \nu_3 $ (neutrino "horizontal conjugation") --- that, as shown in this note, characterizes the phenomenologically favored form (21) of neutrino mixing matrix --- may play an important role in Nature because of the absence for neutrinos of electromagnetic and strong interactions. Otherwise, these interactions could largely suppress such a fragile, discrete horizontal symmetry that, in contrast to the Standard Model gauge interactions, does not treat equally three fermion generations. This may be also the reason, why the quark mixing matrix does not involve large mixings, in contrast to the lepton mixing matrix ({\it i.e}, neutrino mixing matrix in the flavor representation) observed by means of neutrino oscillations.

\vspace{0.3cm}

\ni {\bf 5. Bimaximal mixing matrix as an approximation}

\vspace{0.3cm}

We know from experiments for solar $\nu_e$'s that the bilarge mixing matrix $U$ given in Eq. (21) is not bimaximal, as $\theta_{12} \sim 33^\circ < 45^\circ $, and so

\begin{equation}
c_{12} \sim 0.84 > \frac{1}{\sqrt2} > s_{12} \sim 0.54 \,.
\end{equation}

\ni But, since both values $c_{12}$ and $s_{12}$ are still large and not very distant from $1/\sqrt2 \simeq 0.71$, one may ask the question, if and to what extent the approximation $c_{12} \simeq 1/\sqrt2 \simeq s_{12}$ may work, leading through Eq. (21) to the approximate bimaximal form for the mixing matrix:

\begin{equation}
U \simeq \left( \begin{array}{rrc} 1/\sqrt2 & 1/\sqrt2 & 0 \\ -1/2 & 1/2 & 1/\sqrt2 \\ 1/2 & -1/2 & 1/\sqrt2  \end{array} \right) \,.
\end{equation}

For such an approximate form of $U$, the unitary transformation (20) implies

\begin{eqnarray}
\nu_e & \simeq & \frac{1}{\sqrt2}(\nu_1 + \nu_2)\,, \nonumber \\ 
\nu_\mu & \simeq & \!\!-\frac{1}{\sqrt2} \frac{1}{\sqrt2}(\nu_1 - \nu_2) + \frac{1}{\sqrt2} \nu_3 \,, \nonumber \\ 
\nu_\tau & \simeq & \frac{1}{\sqrt2} \frac{1}{\sqrt2}(\nu_1 - \nu_2) + \frac{1}{\sqrt2} \nu_3 
\end{eqnarray}

\ni and

\begin{eqnarray}
\nu_1 & \simeq & \frac{1}{\sqrt2} \nu_e -  \frac{1}{\sqrt2} \frac{1}{\sqrt2} (\nu_\mu - \nu_\tau) \,, \nonumber \\ 
\nu_2 & \simeq & \frac{1}{\sqrt2} \nu_e + \frac{1}{\sqrt2} \frac{1}{\sqrt2} (\nu_\mu - \nu_\tau)  \,, \nonumber \\ 
\nu_3 & \simeq & \frac{1}{\sqrt2} (\nu_\mu + \nu_\tau) \,.
\end{eqnarray}

\ni It is easy to see from Eqs. (54) and (55) that now, beside the previous (strict) symmetry (25), where 

$$
\nu_1,\nu_2,\nu_3 \rightarrow -\nu_1,-\nu_2,\nu_3 \;\;\;{\rm induces} \;\;\;\nu_e,\nu_\mu,\nu_\tau \rightarrow -\nu_e,\nu_\tau, \nu_\mu \,,
$$ 

\ni there exist also two (approximate) symmetries, where

$$
\nu_1,\nu_2,\nu_3 \rightarrow -\nu_2,-\nu_1,-\nu_3 \;\;\;{\rm induces} \;\;\;\nu_e,\nu_\mu,\nu_\tau \rightarrow -\nu_e,-\nu_\tau, -\nu_\mu 
$$ 

\ni and

$$
\nu_1,\nu_2,\nu_3 \rightarrow \nu_2,\nu_1,-\nu_3 \;\;\;{\rm induces} \;\;\;\nu_e,\nu_\mu,\nu_\tau \rightarrow \nu_e,-\nu_\mu, -\nu_\tau \,.
$$ 

\ni Both are excluded if it is considered that $c_{12} \neq s_{12}$ distinctly. If it is accepted that $c_{12} \simeq s_{12}$, then --- in addition to the relation (25) --- two new relations

\begin{equation}
\left( \begin{array}{rrr} -1 & 0 & 0 \\  0 & 0 & -1 \\  0 & -1 & 0 \end{array}\right) U \left( \begin{array}{rrr} 0 & -1 & 0 \\ -1 & 0 & 0 \\ 0 & 0 & -1 \end{array}\right) \simeq U 
\end{equation}

\ni and

\begin{equation}
\left( \begin{array}{rrr} 1 & 0 & 0 \\  0 & -1 & 0 \\  0 & 0 & -1 \end{array}\right) U \left( \begin{array}{rrr} 0 & 1 & 0 \\ 1 & 0 & 0 \\ 0 & 0 & -1 \end{array}\right) \simeq U 
\end{equation}

\ni follow, respectively, expressing two new (approximate) invariances of the (approximate) form (53) of $U$. These two (approximate) symmetries introduce the difference between the (approximate) bimaximal form (53) of $U$ and its (strict) monomaximal form (21) where $c_{12} \neq s_{12}$ distinctly, because they work only for the former.

Let us denote our Hermitian and real $3\times 3$ matrices transforming the mixing matrix $U$ in the relations (56), (57) and (25) as

\begin{equation}
\widehat{\varphi}_1 \equiv \left( \begin{array}{rrr} -1 & 0 & 0 \\  0 & 0 & -1 \\  0 & -1 & 0 \end{array} \right) \;,\; \widehat{\varphi}_2 \equiv \left( \begin{array}{rrr} 1 & 0 & 0 \\ 0 & -1 & 0 \\ 0 & 0 & -1 \end{array}\right) \;,\; \widehat{\varphi}_3 \equiv \left( \begin{array}{rrr} -1 & 0 & 0 \\ 0 & 0 & 1 \\ 0 & 1 & 0 \end{array}\right)  
\end{equation}

\ni and

\begin{equation}
\widehat{\mu}_1 \equiv \left( \begin{array}{rrr} 0 & -1 & 0 \\ -1 & 0 & 0 \\ 0 & 0 & -1 \end{array}\right)  \;,\; \widehat{\mu}_2 \equiv \left( \begin{array}{rrr} 0 & 1 & 0 \\ 1 & 0 & 0 \\ 0 & 0 & -1 \end{array}\right) \;,\;\widehat{\mu}_3 \equiv \left( \begin{array}{rrr} -1 & 0 & 0 \\ 0 & -1 & 0 \\ 0 & 0 & 1 \end{array}\right) \;.
\end{equation}

\ni Then, with $i = 1,2,3$ the symmetries (56), (57) and (25) of the (approximate) bimaximal mixing matrix $U$ given in Eq. (53) may be expressed in three ways:

\begin{equation}
\widehat{\varphi}_i U \widehat{\mu}_i  = U \;\;{\rm or}\;\;  U\widehat{\mu}_i = \widehat{\varphi}_i U \;{\rm or}\; \widehat{\varphi}_i = U \widehat{\mu}_i U^\dagger \;,
\end{equation}

\ni where for $i = 1,2$ the equality is (only) approximate [while for $i = 3$ it is exact with $U$ as given in Eq. (21)]. Note that Tr~$\widehat{\varphi}_i = -1$ and Tr~$\widehat{\mu}_i = -1$.

From the definitions (58) and (59) we can readily show that with $i,j = 1,2,3$

\begin{equation}
\widehat{\varphi}^2_i = \widehat{1} \;\,,\;\, \widehat{\varphi}_1 \widehat{\varphi}_2 = \widehat{\varphi}_3\;\,{\rm (cyclic)} \;\,,\;\, \widehat{\varphi}_i \widehat{\varphi}_j = \widehat{\varphi}_j \widehat{\varphi}_i \;\,,\;\, \widehat{\varphi}_1+\widehat{\varphi}_2+\widehat{\varphi}_3 =
-\widehat{1}
\end{equation}

\ni and

\begin{equation}
\widehat{\mu}^2_i = \widehat{1} \,\;\,,\;\,\, \widehat{\mu}_1 \widehat{\mu}_2 = \widehat{\mu}_3\;\,{\rm (cyclic)} \;\,,\;\,\, \widehat{\mu}_i \widehat{\mu}_j = \widehat{\mu}_j \widehat{\mu}_i \,\;\,,\;\,\, \widehat{\mu}_1+\widehat{\mu}_2+\widehat{\mu}_3 = -\widehat{1}
\end{equation}

\ni (but in general $\widehat{\varphi}_i \widehat{\mu}_j \neq \widehat{\mu}_j \widehat{\varphi}_i$). Hence, we get in a more compact notation

\begin{equation}
\{\widehat{\varphi}_i \,,\, \widehat{\varphi}_j \} = 2\delta_{i j} \widehat{1} + 2 \sum_k |\varepsilon_{i j k}| \,\widehat{\varphi}_k \;\,,\;\, [\widehat{\varphi}_i \,,\, \widehat{\varphi}_j] = 0
\end{equation}

\ni and

\begin{equation}
\{\widehat{\mu}_i \,,\, \widehat{\mu}_j \} = 2\delta_{i j} \widehat{1} + 2 \sum_k |\varepsilon_{i j k}| \,\widehat{\mu}_k \,\;\,,\;\,\, [\widehat{\mu}_i\,,\, \widehat{\mu}_j] = 0
\end{equation}

\ni [but $\{\widehat{\varphi}_i\,,\, \widehat{\mu}_j\} \neq 0 $ and in general $[\widehat{\varphi}_i\,,\, \widehat{\mu}_j] \neq 0 $, {\it cf.} Eq. (72)]. We can also write for $U$ given in Eq. (53)

\begin{equation}
\left( \delta_{i j}\widehat{1} + \sum_k |\varepsilon_{i j k}|\widehat{\varphi}_k \right) U = \widehat{\varphi}_i U \widehat{\mu}_j = U \left(\delta_{i j} \widehat{1} + \sum_k |\varepsilon_{i j k}| \,\widehat{\mu}_k \right) \,\;,
\end{equation}

\ni where for $j = 1,2$ and $i = 1,2$, respectively, the first and second equality is only approximate [while for $j = 3$ and $i = 3$ it is exact with $U$ as given in Eq. (21)]. Note that here  $\left(|\varepsilon_{i j 1}|\right) = \widehat{\lambda}_6$, $\left(|\varepsilon_{i j 2}|\right) = \widehat{\lambda}_4$ and $\left( |\varepsilon_{i j 3}|\right) = \widehat{\lambda}_1$.

It is easy to see that the effective mass matrix $M^{\rm eff} = U\,{\rm diag}(m_1,m_2,m_3) U^\dagger $ with $U$ in the form (53) reveals for $i = 1,2,3 $ the symmetries

\begin{equation}
\widehat{\varphi}_i M^{\rm eff} \widehat{\varphi}_i = M^{\rm eff}\;\;{\rm or}\;\;M^{\rm eff} \widehat{\varphi}_i = \widehat{\varphi}_i M^{\rm eff}  \,,
\end{equation}

\vspace{0.2cm}

\ni where for $i = 1,2$ the equality is (only) approximately valid, provided we can accept beside the approximation $c_{12} \simeq s_{12}$ {\it also} $m_1 \simeq m_2$ {\it i.e.}, $\delta \simeq 0$ [while for $i = 3$ it is exact with $U$ as given in Eq. (21)]. In the case of $\delta \simeq 0$, the values of $c_{12}$ and $s_{12}$ become irrelevant in $M^{\rm eff}$ [{\it cf.} Eqs. (50)].

The matrices $\widehat{\varphi}_i$ and $\widehat{\mu}_i\;(i = 1,2,3)$ as defined in Eqs. (58) and (59) may be used as bases for $3\times 3$ symmetric block matrices of the types

$$
\left( \begin{array}{rrr} A & 0 & 0 \\ 0 & B & C \\ 0 & C & B \end{array}\right) \;\;{\rm and}\;\; \left( \begin{array}{rrr} D & E & 0 \\  E & D & 0 \\ 0 & 0 & F \end{array}\right) \;,
$$

\vspace{0.2cm}

\ni respectively ($\widehat{1}$ is not needed in these bases because of the form of constraints $\widehat{\varphi}_1+\widehat{\varphi}_2+\widehat{\varphi}_3 = -\widehat{1}$ and $\widehat{\mu}_1+\widehat{\mu}_2+\widehat{\mu}_3 = -\widehat{1}$). The sets of such matrices form two Abelian groups with respect to matrix multiplication, if the inverse of their four blocks exists. They are isomorphic, as they are related through the unitary transformation generated by the bimaximal mixing matrix $U$ given by the rhs of Eq. (53), $(I) = U (II) U^\dagger $, where $(I)$ and $(II)$ symbolize the sets of matrices of the first and second type. The group character of these sets is reflected in the group relations $\widehat{\varphi}_1 \widehat{\varphi}_2 = \widehat{\varphi}_3\;{\rm (cyclic)}$ and $\widehat{\mu}_1 \widehat{\mu}_2 = \widehat{\mu}_3\;{\rm (cyclic)}$ for their bases [{\it cf.} Eqs. (61) and (62)], while their isomorphism corresponds to the unitary transformation $\widehat{\varphi}_i = U \widehat{\mu}_i U^\dagger \;\;(i = 1,2,3)$ between both bases [{\it cf.} Eqs. (60)]. These two groups are, of course, subgroups of the group of all $3\times 3$ matrices that can be spun by the basis consisting of $\widehat{1}$ and the Gell-Mann matrices $\widehat{\lambda}_a \;(a = 1,2,\ldots,8)$. 

It is interesting to note that the neutrino effective mass matrix $M^{\rm eff}$ is of the form [{\it cf.} Eqs. (50)] belonging to the set $(I)$ for $d = 0$:

$$
\left( \begin{array}{rrr} a & d & -d \\ d & b & c \\ -d & c & b \end{array}\right) \;.
$$

\ni This form commutes exactly with $\widehat{\varphi}_3$, while with $ \widehat{\varphi}_1$ and $\widehat{\varphi}_2$ (only) approximately, provided $d \equiv \delta s/\sqrt2 \simeq 0$ and so $ \delta  \simeq 0$ {\it i.e.}, $m_1 \simeq m_2$. Here, $\widehat{\varphi}_3 = U \widehat{\mu}_3 U^\dagger \equiv U P^{(H)} U^\dagger$ is the unitary transform of "horizontal parity"~with $U$ as given in Eq. (21). In this way, $\widehat{\varphi}_3 M^{\rm eff} \widehat{\varphi}_3 = U P^{(H)} {\rm diag} (m_1,m_2,m_3) P^{(H)}U^\dagger = M^{\rm eff}$. In terms of $ \widehat{\varphi}_1,\, \widehat{\varphi}_2,\, \widehat{\varphi}_3$ and $\widehat{\lambda}_1,\,\widehat{\lambda}_4$ with $(\widehat{\varphi}_1+\widehat{\varphi}_2)(\widehat{\varphi}_1-\widehat{\varphi}_2)=0$ and $(\widehat{\varphi}_1+\widehat{\varphi}_2)(\widehat{\lambda}_1-\widehat{\lambda}_4)=0$ we can write

\begin{equation}
M^{\rm eff} = -\left(\stackrel{0}{m}+ \frac{1}{2} \Delta \right) \left( \widehat{\varphi}_1 + \widehat{\varphi}_2 \right) - \stackrel{0}{m} \widehat{\varphi}_3 + \frac{1}{2}\delta\,c \left( \widehat{\varphi}_1 - \widehat{\varphi}_2 \right) + \frac{1}{\sqrt2} \delta s  \left( \widehat{\lambda}_1 - \widehat{\lambda}_4 \right)\,,
\end{equation}

\ni where $\widehat{\varphi}_3 \widehat{\lambda}_{1,4}\! = \!- \widehat{\lambda}_{4,1} \widehat{\varphi}_3$ makes $\widehat{\varphi}_3$ commute with $\widehat{\lambda}_1 - \widehat{\lambda}_4$, while $ \widehat{\varphi}_1 \widehat{\lambda}_{1,4} = \widehat{\lambda}_{4,1} \widehat{\varphi}_1$ and $\widehat{\varphi}_2 \widehat{\lambda}_{1,4} = - \widehat{\lambda}_{1,4}\widehat{\varphi}_2$ imply that $\widehat{\varphi}_1$ and $\widehat{\varphi}_2$ anticommute with $\widehat{\lambda}_1 - \widehat{\lambda}_4$. Also note that $ U^\dagger \left[\frac{1}{2} c \left( \widehat{\varphi}_1 - \widehat{\varphi}_2 \right) + \frac{1}{\sqrt2} s  \left( \widehat{\lambda}_1 - \widehat{\lambda}_4 \right)\right] U = - \widehat{\lambda}_3$ for any $ c \equiv c^2_{12}-s^2_{12}$ and $ s \equiv 2c_{12} s_{12}$ [{\it cf.} Eqs. (43)], where $U$ is given as in Eq. (21). If $c_{12} = 1/\sqrt2 = s_{12}$, then $c = 0$ and $s = 1$.

In connection with the formula (67) we wonder, if the $3\times 3$ matrices $\widehat{\varphi}_i$ and $\widehat{\mu}_i\; (i = 1,2,3)$ may help us to find the desired dynamical variables solving hopefully the basic pro\-blem of fermion masses. In such a case there may appear a more or less instructive analogy with Pauli matrices that have led to Dirac matrices solving the problem of fermion spins.

In terms of the Gell-Mann $3\times 3$ matrices $\widehat{\lambda}_a\;(a = 1,2,\ldots,8)$ and their basic combinations $\widehat{e}_{\alpha \beta}\; (\alpha,\beta = e,\mu,\tau)$ presented in Eqs. (49) we obtain

\begin{equation}
\widehat{\varphi}_1 = -\widehat{\lambda}_6 - \widehat{e}_{e e} \;,\, \widehat{\varphi}_2 = -\widehat{1} + 2\widehat{e}_{e e} \;,\; \widehat{\varphi}_3 = \widehat{\lambda}_6 - \widehat{e}_{e e}
\end{equation}

\ni and

\begin{equation}
\widehat{\mu}_1 = -\widehat{\lambda}_1 - \widehat{e}_{\tau \tau} \;,\; \widehat{\mu}_2 = \widehat{\lambda}_1 - \widehat{e}_{\tau \tau} \;,\, \widehat{\mu}_3 = -\widehat{1} + 2\widehat{e}_{\tau \tau} \;,
\end{equation}

\ni where

\begin{equation}
\widehat{e}_{e e} = \left( \begin{array}{rrr} 1 & 0 & 0 \\ 0 & 0 & 0 \\ 0 & 0 & 0 \end{array}\right) =
\frac{1}{3}\widehat{1} + \frac{1}{2}\left( \widehat{\lambda}_3 + \frac{1}{\sqrt3} \widehat{\lambda}_8 \right) \;,\; \widehat{e}_{\tau \tau} = \left( \begin{array}{rrr} 0 & 0 & 0 \\ 0 & 0 & 0 \\ 0 & 0 & 1 \end{array}\right) = \frac{1}{3}\widehat{1} - \frac{1}{\sqrt3} \widehat{\lambda}_8  \;. 
\end{equation}

\ni Due to Eqs. (33) and (36) we can write for $i = 3$

\begin{equation}
\widehat{\mu}_3 =  e^{i \widehat{\lambda}_2 \pi} \;,\; \widehat{\varphi}_3 = e^{i \widehat{\lambda}_7 \pi/4} e^{i \widehat{\lambda}_2 \pi} e^{-i\widehat{\lambda}_7 \pi/4} = U e^{i \widehat{\lambda}_2 \pi} U^\dagger
\end{equation}

\ni with the mixing matrix $U$ as presented in Eq. (30). For $i = 3$ it is not necessary to make in $U$ the approximation $\theta_{12} \simeq 45^\circ $ (here, this mixing angle has its actual value $\theta_{12} \sim 33^\circ $).

Since the matrices $\widehat{\varphi}_i$ and $\widehat{\mu}_i \;(i = 1,2,3)$ are Hermitian and real, their commutators are antiHermitian and real, and thus, can be expressed as combinations of three imaginary matrices $\widehat{\lambda}_2\,,\,\widehat{\lambda}_5$ and $\widehat{\lambda}_7 $ of eight Gell-Mann $3\times 3$ matrices $\widehat{\lambda}_a\;(a = 1,2,\ldots,8)$ which all are Hermitian. In fact, we find

\begin{equation}
\left(\left[\widehat{\varphi}_i,\widehat{\mu}_j \right] \right) =  i\left( \begin{array}{ccc} \widehat{\lambda}_2-\widehat{\lambda}_5+\widehat{\lambda}_7 & -\widehat{\lambda}_2+\widehat{\lambda}_5+\widehat{\lambda}_7 & -2\widehat{\lambda}_7 \\ -2\widehat{\lambda}_2 & 2\widehat{\lambda}_2 & 0 \\ \widehat{\lambda}_2+\widehat{\lambda}_5-\widehat{\lambda}_7 & -\widehat{\lambda}_2-\widehat{\lambda}_5-\widehat{\lambda}_7 & 2\widehat{\lambda}_7 \end{array}\right) \,.
\end{equation}

\vspace{0.2cm}

\ni Hence, $\sum_i \left[ \widehat{\varphi}_i,\widehat{\mu}_j \right] = 0$ and $\sum_j \left[ \widehat{\varphi}_i , \widehat{\mu}_j \right] = 0$, what is consistent with two constraints $\sum_i  \widehat{\varphi}_i = -\widehat{1} $ and $\sum_j \widehat{\mu}_j = -\widehat{1} $. The imaginary matrices $\widehat{\lambda}_2,\widehat{\lambda}_5$ and $\widehat{\lambda}_7$ are absent from $ \widehat{\varphi}_i$ and $\widehat{\mu}_i$ that are combinations of $\widehat{1}, \widehat{\lambda}_1,\widehat{\lambda}_3,\widehat{\lambda}_8$ and $\widehat{\lambda}_6$ [{\it cf.} Eqs. (68)--(70)].

Analogically, it is easy to evaluate the anticommutators $\left\{\widehat{\varphi}_i, \widehat{\mu}_j \right\}$  that can be expressed as combinations of $\widehat{1}$ and five real matrices $\widehat{\lambda}_1,\widehat{\lambda}_3,\widehat{\lambda}_4, \widehat{\lambda}_6$ and $\widehat{\lambda}_8$ of eight $\widehat{\lambda}_a\;(a = 1,2,\ldots,8) $. None of these anticommutators is zero. 

Note that the matrices $ \widehat{\varphi}_i$ and $\widehat{\mu}_i$ defined in Eqs. (58) and (59) have the following block structure

\vspace{-0.3cm}

\begin{equation}
\widehat{\varphi}_1= \left( \begin{array}{rccc} -1\;\; & 0\!\! & ~& \!\!0 \\ 0\;\; & \!\! & ~ & \!\! \\  \;\; &  \!\! & \!-\widehat{\sigma}^P_1 & \!\!  \\ 0\;\; & \!\! & ~ & \!\! \end{array}\right)\,,\, \widehat{\varphi}_2= \left( \begin{array}{rccc} 1\;\; & 0\!\! & ~& \!\!0 \\ 0\;\; & \!\! & ~ & \!\! \\ \;\; & \!\! & \!-\widehat{1}^P & \!\!  \\ 0\;\; & \!\! & ~ & \!\! \end{array}\right)\;,\;\widehat{\varphi}_3 = \left( \begin{array}{rccc} -1\;\; & 0\!\! & ~& \!\!0 \\ 0\;\; & \!\! & ~ & \!\! \\  \;\; &  \!\! & \widehat{\sigma}^P_1 & \!\!  \\ 0\;\; & \!\! & ~ & \!\! \end{array}\right) 
\end{equation}

\ni and 

\vspace{-0.3cm}

\begin{equation}
\widehat{\mu}_1= \left( \begin{array}{cccr} \! & \!\! & \! & \;\;0 \\  \!\! & \!-\widehat{\sigma}^P_1 & \!\! & \;\; \\ \!\! & \! & \!\! & \;\;0 \\ 0\!\! & \! & \!\!0 & \;-1 \end{array}\right)\,,\, 
\widehat{\mu}_2 = \left( \begin{array}{cccr} \!\! & \! & \!\! & \;0 \\ \!\! & \widehat{\sigma}^P_1 & \!\! & \; \\  \!\! & \! & \!\! & \;\;0 \\ 0\!\! & \!& \!\!0 & \;-1 \end{array}\right)\,,\,
\widehat{\mu}_3= \left( \begin{array}{cccr} \!\! & \! & \!\! & \;\;0 \\ \!\! & \!-\widehat{1}^P & \!\! & \;\; \\ \!\! & \! & \!\! & \;\;0 \\ 0\!\! & \! & \!\!0 & \;\;1 \end{array}\right) \;,
\end{equation}

\ni where

\vspace{-0.2cm}

\begin{equation}
\widehat{1}^P =  \left( \begin{array}{rr} 1 & 0 \\ 0 & 1 \end{array}\right) \;,\; \widehat{\sigma}^P_1 = \left( \begin{array}{rr} 0 & 1 \\ 1 & 0 \end{array}\right) \;.
\end{equation}

\vspace{0.3cm}

\ni {\bf 6. Conclusions}

\vspace{0.3cm}

We have introduced in Section 5 two algebras of commuting Hermitian $3\times 3$ matrices $ \widehat{\varphi}_1,\widehat{\varphi}_2,\widehat{\varphi}_3$ and $\widehat{\mu}_1, \widehat{\mu}_2,\widehat{\mu}_3$, satisfying the group relations $\widehat{\varphi}_1 \widehat{\varphi}_2 = \widehat{\varphi}_3\;{\rm (cyclic)}$ and $\widehat{\mu}_1 \widehat{\mu}_2 = \widehat{\mu}_3\;{\rm (cyclic)}$ as well as the constraints $ \widehat{\varphi}_1 +\widehat{\varphi}_2+\widehat{\varphi}_3 = -\widehat{1}$ and $\widehat{\mu}_1 +\widehat{\mu}_2+\widehat{\mu}_3 = -\widehat{1}$. These two algebras are isomorphic, as being related by the unitary transformation $\widehat{\varphi}_i = U\,\widehat{\mu}_i \,U^\dagger \;(i = 1,2,3)$, where $U$ is the neutrino mixing matrix of the bilarge form (21), phenomenologically favored at present, which for $i = 1,2$ is approximated to the nearly bimaximal form (53) with $c_{12} \simeq 1/\sqrt2 \simeq s_{12}$ ({\it i.e.}, with $\theta_{12} \simeq 45^\circ $, while the actual experimental estimate is $\theta_{12} \sim 33^\circ $). For $i = 3$ the unitary transformation is exact, thus the resulting {\it invariance} $\widehat{\varphi}_3 U\widehat{\mu}_3 = U$ is also exact. It {\it characterizes} the monomaximal form (21) of neutrino mixing matrix $U$ for any $c_{12}$ and $s_{12}$. The resulting {\it approximate invariances} $\widehat{\varphi}_i U\widehat{\mu}_i = U$ for $i = 1,2$, if accepted as working, {\it suggest} that this monomaximal form of $U$ is, in fact, nearly bimaximal. Of course, the closer the experimetal estimate of $\theta_{12}$ is to $45^\circ $, the better is this conclusion.

We have called the transformation $ U' = \widehat{\varphi}_3 U\widehat{\mu}_3$ providing the exact invariance $U' = U$ the neutrino "horizontal conjugation", while its matrix $P^{(H)} \equiv \widehat{\mu}_3 = \exp(i \widehat{\lambda}_2 \pi)$ --- the neutrino "horizontal parity". The latter is a unitary rotation by the angle $2\pi $ around the "horizontal"~2-axis. Such a conjugation implies the transformation

\begin{equation}
\left( \begin{array}{r} \nu'_1 \\ \nu'_2 \\ \nu'_3 \end{array} \right) = P^{(H)} \left( \begin{array}{r} \nu_1 \\ \nu_2 \\ \nu_3 \end{array} \right) = \left( \begin{array}{r} -\nu_1 \\ -\nu_2 \\ \nu_3 \end{array} \right)
\end{equation}

\vspace{0.2cm}

\ni for mass neutrinos, inducing simultaneously the transformation

\begin{equation}
\left( \begin{array}{r} \nu'_e \\ \nu'_\mu \\ \nu'_\tau \end{array} \right) =U P^{(H)} U^\dagger \left( \begin{array}{r} \nu_e \\ \nu_\mu \\ \nu_\tau \end{array} \right) = \left( \begin{array}{r} -\nu_e \\ \nu_\tau \\ \nu_\mu \end{array} \right)
\end{equation}

\vspace{0.2cm}

\ni for flavor neutrinos, since $\nu'_\alpha = \sum_i U_{\alpha i} \nu'_i$ and $\nu_i = \sum_\alpha U^*_{\alpha i} \nu_\alpha$. Thus, the "horizontal parity"~displays the {\it covariance} ${P^{(H)}}' = U P^{(H)} U^\dagger $ with ${P^{(H)}}' \equiv \widehat{\varphi}_3$. Here, the mass neutrinos $\nu_1 \,,\, \nu_2 \,,\, \nu_3$ get the "horizontal parity"~equal to -1, -1, 1, respectively, while the flavor neutrinos $\nu_e \,,\, \nu_\mu \,,\, \nu_\tau $, except for $\nu_e $, mix the "horizontal parity".

The corresponding transformation for the neutrino effective mass matrix $M^{\rm eff} = U\,{\rm diag}(m_1,m_2,m_3)\, U^\dagger $ reads ${M^{\rm eff}}' = \widehat{\varphi}_3 M^{\rm eff} \widehat{\varphi}_3$, implying the exact invariance ${M^{\rm eff }}' =  M^{\rm eff}$ because of the relations $\widehat{\varphi}_3 U =  U\widehat{\mu}_3$ and $ \widehat{\mu}_3 \,{\rm diag}(m_1,m_2,m_3)\, \widehat{\mu}_3 = {\rm diag}(m_1,m_2,m_3) $. The invariances $ \widehat{\varphi}_i M^{\rm eff} \widehat{\varphi}_i = M^{\rm eff}$ for $i = 1,2$ are (only) approximately valid, provided we can accept beside the approximation $c_{12} \simeq s_{12}$ {\it also} $m_1 \simeq m_2$.

Thus, finally, we can conclude that --- as a result of there being the "horizontal parity"~covariant in neutrino mixings --- three active neutrinos may develop {\it in a natural way} the familiar bilarge form of mixing matrix, favored at present phenomenologically. The suggestion that the resulting monomaximal form is, in fact, bilarge (approximately bimaximal) comes from the group character of relations $ \widehat{\mu}_1 \widehat{\mu}_2 =  \widehat{\mu}_3$ (cyclic), where the "horizontal parity" $P^{(H)} \equiv \widehat{\mu}_3$ (generating the symmetry  under the "horizontal conjugation", $\widehat{\varphi}_3 U \widehat{\mu}_3 = U$) is embedded. Then, the symmetries
$\widehat{\varphi}_i U \widehat{\mu}_i = U$ or $\widehat{\varphi}_i = U \widehat{\mu}_i U^\dagger$ with $ \widehat{\varphi}_1 \widehat{\varphi}_2 =  \widehat{\varphi}_3$ (cyclic) hold also for $i = 1,2$, if $U$ --- monomaximal with some $c_{12}$ and $s_{12}$ --- is approximated to bimaximal form with $c_{12}\rightarrow 1/\sqrt2 \leftarrow s_{12}$, in order to include also $ \widehat{\varphi}_1,\widehat{\varphi}_2$ and $\widehat{\mu}_1,\widehat{\mu}_2$ in describing potential symmetries of $U$. Obviously, the smaller are the experimentally estimated differences $c_{12} - 1/\sqrt2 $ and $s_{12} - 1/\sqrt2 $, the better are the approximated symmetries for $i = 1,2$.

\vfill\eject

~~~~
\vspace{0.5cm}

{\centerline{\bf References}}

\vspace{0.5cm}

{\everypar={\hangindent=0.6truecm}
\parindent=0pt\frenchspacing

{\everypar={\hangindent=0.6truecm}
\parindent=0pt\frenchspacing

[1]~W. Kr\'{o}likowski, {\tt hep--ph/0301161}.

\vspace{0.2cm}

[2]~S. Fukuda {\it et al.} (SuperKamiokande Collaboration), {\it Phys. Rev. Lett.} {\bf 85}, 3999 (2000).

\vspace{0.2cm}

[3]~K. Eguchi {\it et al.} (KamLAND Collaboration), {\it Phys. Rev. Lett.} {\bf 90}, 021802 (2003).

\vspace{0.2cm}

[4]~V. Barger and D. Marfatia, {\tt hep--ph/0212126}.

\vspace{0.2cm}

[5]~G.L. Fogli {\it et al.}, {\tt hep--ph/0212127}.

\vspace{0.2cm}

[6]~M. Maltoni, T. Schwetz and J.W.F. Valle, {\tt hep--ph/0212129}.

\vspace{0.2cm}

[7]~A. Bandyopadhyay {\it et al.}, {\tt hep--ph/0212146v2}.

\vspace{0.2cm}

[8]~J.N.~Bahcall, M.C.~Gonzalez--Garcia and C. Pe\~{n}a--Garay, {\tt hep--ph/0212147v2}.

\vspace{0.2cm}

[9]~M. Appolonio {\it et al.} (Chooz Collaboration), {\tt hep--ex/0301017}.

\vspace{0.2cm}

[10]~G. Mills, {\it Nucl. Phys. Proc. Suppl.} {\bf 91}, 198 (2001).

\vspace{0.2cm}

[11]~C. Giunti, {\tt hep--ph/0302173}.

\vfill\eject

\end{document}